# Enhanced thermopower upon Ni substitution in $Na_{0.75}CoO_2$

**N. Gayathri*, A.Bharathi and Y.Hariharan**
Materials Science Division, Indira Gandhi Centre for Atomic Research, Kalpakkam, 603102
*Corresponding author: E-mail: gayathri@igcar.ernet.in

***Abstract***

*Thermopower measurements have been carried out on the Ni substituted samples of $Na_{0.75}CoO_2$ in the temperature range 4.2K to 300K. The room temperature TEP increases by 20μV/K even with 1% Ni substitution and systematically increases with increasing Ni content upto 5%. The increase in TEP is accompanied by a decrease in ρ thus increasing the ratio of $S^2/\rho$ on Ni substitution. At low T, the TEP shows an anomaly in the substituted samples, showing a peak at T~ 20K.*

## INTRODUCTION

The recent extensive studies on $Na_yCoO_2$ have revealed a range of interesting and intriguing properties of this system. This system shows two metallic phases separated by an insulating phase at y=0.5 [1]. The lower y compounds in its hydrated form show superconductivity with a maximum $T_C$ ~ 5K [2]. The unhydrated compound is interesting in its own right owing to the exceptionally high thermopower over the range $0.5 \le y \le 0.9$, which, unusually, accompanies low thermal conductivity and low resistivity [3]. Bulk antiferromagnetic order has been observed below 20K for $0.7 < y < 0.9$, the compound still being metallic [4]. A general understanding has been reached that the high thermopower arises because of the spin entropy gained in the transport of an electron from low spin $Co^{3+}$ state to a low spin $Co^{4+}$ state [5,6]. The ratio of $Co^{3+}/Co^{4+}$ is thus important in determining the ground state properties of the $Na_yCoO_2$ and hence any substitution that changes this ratio will affect the resistivity as well as the magnitude of the thermopower. The effect of Ni substitution on the resistivity behaviour of this system has been studied in detail [7]. The system investigated $Na_{0.75}Co_{1-x}Ni_xO_2$ ($0 \le x \le 0.15$) shows a metal-insulator transition as a function of temperature as the temperature is lowered, and the $T_{MIT}$ increases as the Ni content increases. The nature of the metallic state of the substituted samples are significantly different from the unsubstituted sample with the resistivity having a $T^2$ temperature dependence compared to the pristine sample showing a sublinear temperature dependence. The metal-insulator transition has also been confirmed by IR spectroscopy [7]. The Fano asymmetry parameter extracted from the IR active in-plane Co-O mode agrees with the resistivity behaviour of the system as a function of Ni substitution as well as temperature. This drastic ground state changes on Ni substitution can have significant implication on the thermoelectric property of this material.

In this regard, we have made thermopower measurements on the Ni substituted samples and we present the results here.

## EXPERIMENTAL DETAILS

The samples of $Na_{0.75}Co_{1-x}Ni_xO_2$ ($0 \le x \le 0.15$) were prepared by standard solid state reaction as discussed in Ref [7]. The thermopower measurements on some of the Ni substituted samples were carried out by the dc-technique in an exchange gas cryostat. In the experimental arrangement the sample was sandwiched between two copper blocks using tripod spring arrangement. The ΔT achieved with the help of a manganin heater placed below one of the Cu blocks was measured using a differential (Au:0.07%Fe, Chromel) thermocouple attached to the blocks and referenced to the cold end of the sample (measured by a cernox thermometer). The ΔV was measured using two Cu wires spot welded to the Cu blocks very close to the region of contact with the sample. In a typical measurement, after temperature stabilization, the ΔT heater was switched on and the ΔV and ΔT were recorded simultaneously until the ΔT incremented to ~1K. In this regime a linear behaviour was ensured. The thermopower of the sample is given by $S_{sample} = S_{Cu} - (\Delta V.\text{verses}.\Delta T)_{slope}$. The full data acquisition was done with software written under Labview 6.0.

## RESULTS AND DISCUSSION

Fig.1 shows the S measured at room temperature on all samples presented in this study. The resistivity obtained for each of the samples at room temperature is also plotted in the same figure. From the figure it is evident that the thermopower increases with Ni content and for the x=0.01 and x=0.02 the increase in thermopower is associated with a decreases in resistivity at RT. It is well known that the figure of merit deciding the usefulness of a thermoelectric material is given by $Z=S^2/\rho k$. From the measured values of S and ρ, shown in Fig1(a), $S^2/\rho$ is evaluated and is shown in Fig.1(b). The figure clearly indicates that there is a three fold increase in $S^2/\rho$ suggesting that the usefulness of $Na_yCoO_2$ as a thermoelectric material is enhanced by Ni substitution.

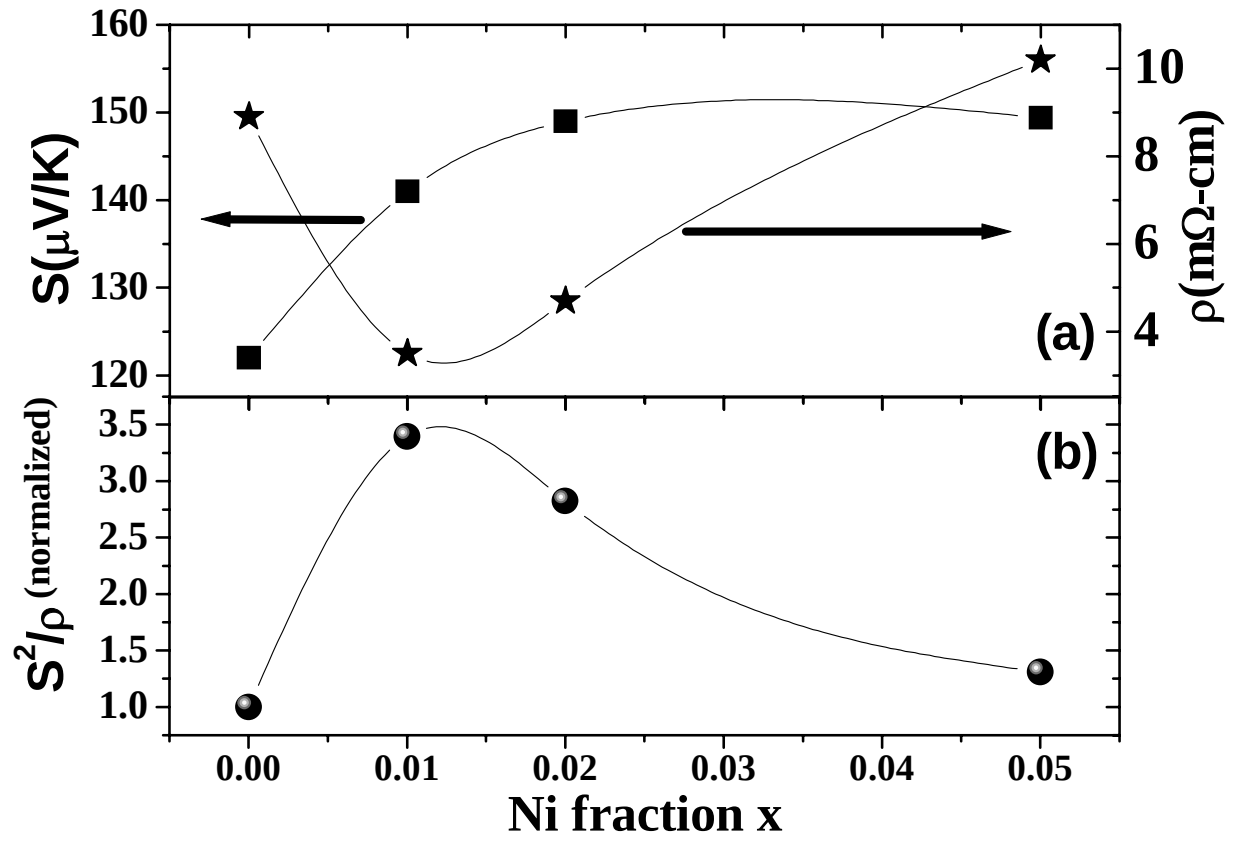

**Fig.1(a)**: Room temperature thermopower S (■) and resistivity ρ (★) versus Ni fraction, solid lines are guides to the eye. **(b)**: Ratio of $S^2/\rho$ versus Ni fraction, normalized to the value in the pristine sample.

Fig.2. shows the temperature dependence of the thermopower normalized to the value at 300K in all the samples. It is evident from the figure that the temperature dependence is similar for all samples, excepting for the low temperature region (< 30K) where a peak is observed in the Ni substituted samples (inset of fig.2). Thus while the absolute value of thermopower increases with Ni substitution, the overall temperature dependence remains unaltered, indicating that the mechanism governing the S(T) is unaltered due to Ni substitution. An indication of the high temperature limit of the thermopower can be obtained by using Heikes formula [6]

$$S(T) \rightarrow \frac{-k_B}{e}\ln\left(\frac{g_3}{g_4}\left(\frac{n}{1-n}\right)\right)$$

where $g_3$, $g_4$ are the degeneracies of the electronic states in $Co^{3+}$ and $Co^{4+}$ ions (low spin) and n is the number of $Co^{4+}$ ions and charge transport by holes is considered. Since Ni enters the lattice in the 4+, low spin state, its degeneracy is the same as that of $Co^{3+}$, whereas the degeneracy of $Co^{4+}$ is 6. With Ni substitution n decreases from value of 0.25 in the unsubstituted sample to 0.25-x in the Ni substituted samples. Inserting these values into the above equation it can easily be verified that S increases from 249μV/K to 274μV/K with 5% Ni substitution. The larger increase observed experimentally could arise due to increase in Hubbard U due to Ni substitution [7] as shown in Ref [8]. We now address the temperature dependence of thermopower shown in Fig.2. It is apparent from the figure that apart from the behaviour of the low temperature peak the temperature dependence remains unaffected due to Ni substitution. This is surprising especially when compared with the temperature dependence of resistivity in these samples, where very large differences have been observed for different Ni substitutions on account of the metal-insulator transitions. It is now well accepted that the thermopower in the metallic pristine cobaltate is due to spin transport. Thus, while resistivity probes the charge dynamics of the systems, thermopower probes the spin dynamics. While Ni substitution brings about drastic changes in the former, it does not alter the latter. The low temperature peak may be reminiscent of phonon drag, but its absence in the pristine sample, suggests that its origin could be due to a change in the density of states at $E_F$ on account of an electronic transition. A step like feature seen in some thermopower measurements in pristine cobaltate has been assigned to the SDW transition observed by magnetization measurements [4]. A similar feature in the thermopower has also been seen in the Cu substituted samples [9].

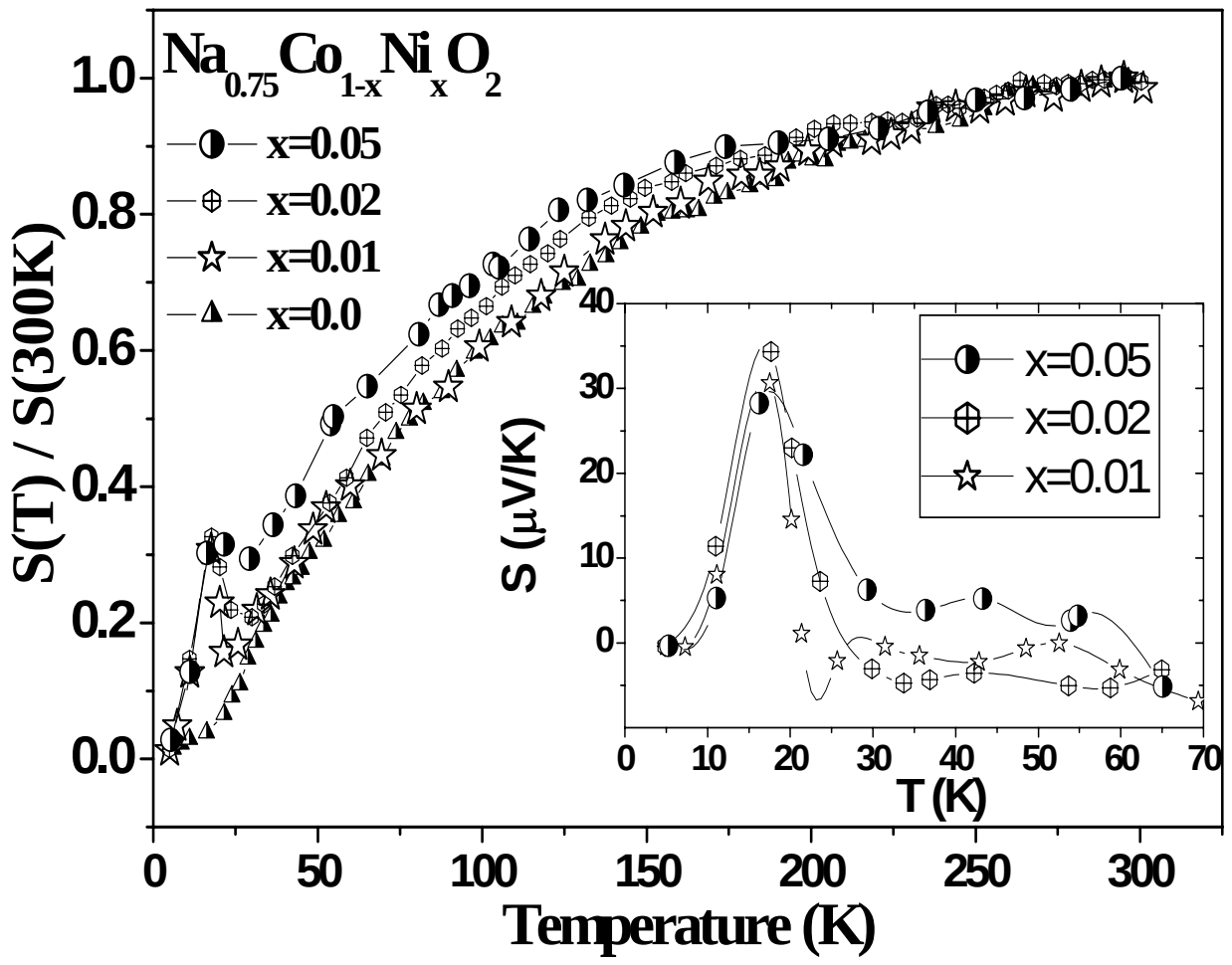

**Fig 2:** Temperature dependence of S normalized to the room temperature value. **Inset:** S at low temperatures after subtraction of a linear background showing the peak in the substituted samples at T~ 20K.